# Deep Network Embedding for Graph Representation Learning in Signed Networks

Xiao Shen and Fu-Lai Chung

*Abstract*—Network embedding has attracted an increasing attention over the past few years. As an effective approach to solve graph mining problems, network embedding aims to learn a low-dimensional feature vector representation for each node of a given network. The vast majority of existing network embedding algorithms, however, are only designed for unsigned networks, and the signed networks containing both positive and negative links, have pretty distinct properties from the unsigned counterpart. In this paper, we propose a deep network embedding model to learn the low-dimensional node vector representations with structural balance preservation for the signed networks. The model employs a semisupervised stacked auto-encoder to reconstruct the adjacency connections of a given signed network. As the adjacency connections are overwhelmingly positive in the real-world signed networks, we impose a larger penalty to make the auto-encoder focus more on reconstructing the scarce negative links than the abundant positive links. In addition, to preserve the structural balance property of signed networks, we design the pairwise constraints to make the positively connected nodes much closer than the negatively connected nodes in the embedding space. Based on the network representations learned by the proposed model, we conduct link sign prediction and community detection in signed networks. Extensive experimental results in real-world datasets demonstrate the superiority of the proposed model over the state-of-the-art network embedding algorithms for graph representation learning in signed networks.

## I. Introduction

NETWORK embedding is an effective method to learn a low-dimensional feature vector representation for each node of a given network, with the goal of preserving the original network structure in the embedding space. Based on the low-dimensional feature vector representations, many graph analytics problems, such as graph visualization, node classification, link prediction, and graph clustering, can all be easily solved. Although several promising network embedding algorithms, such as DeepWalk [1], Line [2], node2vec [3], SDNE [4], DNGR [5]; and DNE-APP [6] have been proposed recently, they are only designed for unsigned networks, without considering the polarities of edges in the signed networks.

A signed network contains both positive and negative links, where the positive links indicate proximity or similarity, while the negative links reflect dissimilarity or distance [7]. Recent studies [8], [9] have shown that the signed networks have properties substantially distinct from unsigned networks. For example, in unsigned networks, the homophily effect [10] and social influence [11] theories suggest that two users directly connected by a link would tend to have similar preferences. However, such theories are not applicable to the signed networks due to the existence of negative links. For instance, in a signed network like Epinions,[1] two negatively connected users would have rather opposite opinions instead of similar preferences. In addition, the transitivity property of unsigned networks which suggests that "the friend of my friend is likely to be my friend" is also not true for the negative links in the signed networks, since both "the enemy of my enemy is my friend" and "the enemy of my enemy is my enemy" can be observed in the signed networks [12]. Due to the substantially distinct properties between the signed networks and unsigned networks, existing network embedding algorithms designed for unsigned networks cannot be directly applied to the signed networks. Thus, it is indeed necessary to design signed network embedding algorithms to capture the specific properties of the signed networks. "Structural balance" is one prevailing social property of the signed networks [13]. The balance theory [14] states that "a network is balanced if and only if all the edges are positive; or all the nodes in the network can be grouped into two clusters where the edges within the same cluster are all positive, while the edges across different clusters are all negative." A weak balance theory [15] was proposed to generalize the original balance theory [14] from two-way clustering to $k$-way clustering. Recently, Cygan *et al.* [16] further extended the structural balance theory as "the nodes connected with positive links should sit closer (i.e., possess higher proximity) than those nodes connected with negative links."

Existing signed network embedding algorithms mostly employ the spectral techniques [7], [17]–[19] to embed the original network into a low-dimensional space spanned by the top-$k$ eigenvectors of the characteristic matrix associated with the given network. It has been shown that such

Manuscript received April 10, 2018; revised July 20, 2018; accepted September 6, 2018. This work was supported in part by the Hong Kong Ph.D. Fellowship Scheme under Grant PF14-11856, and in part by the PolyU CRF Projects (G-YBQH and G-YBVT). This paper was recommended by Associate Editor Y. Zhang. *(Corresponding author: Fu-Lai Chung.)*

The authors are with the Department of Computing, Hong Kong Polytechnic University, Hong Kong (e-mail: shenxiaocam@163.com; cskchung@comp.polyu.edu.hk).

Color versions of one or more of the figures in this paper are available online at http://ieeexplore.ieee.org.

Digital Object Identifier 10.1109/TCYB.2018.2871503

[1]http://www.epinions.com/

spectral methods based on matrix decomposition are with limited representation learning ability to capture the highly nonlinear properties of the complex network structures [5]. In addition, the spectral methods based on eigen value decomposition (EVD) are computationally highly expensive, i.e., even the fastest implementation of EVD requires a superquadratic computational complexity [20]. On the other hand, deep learning techniques have demonstrated the powerful ability to learn more complex and nonlinear feature representations in the areas of computer vision [21], [22], speech recognition [23], and natural language processing (NLP) [24]. Thus, most recently, several promising deep network embedding algorithms [4]–[6], [25] have been proposed to learn graph representations for unsigned networks. However, very little deep network embedding work exists for the signed networks.

In this paper, we propose a deep network embedding with structural balance preservation (DNE-SBP) model to learn graph representations for the signed networks. A stacked autoencoder (SAE) is employed to learn the nonlinear hidden representations, by reconstructing the adjacency matrix of a given signed network. As the real-world signed networks are generally overwhelmingly positive [12], we impose a larger penalty on the reconstruction errors of negative links so as to make the SAE focus more on reconstructing the scarce negative links as compared to the abundant positive links. In addition, it has been shown that the semisupervised learning techniques which incorporate pairwise constraints as the side information can effectively improve the learning performance [4], [6], [25]–[28]. Motivated by this, we further design a semisupervised SAE by incorporating the pairwise constraints to map each positively connected node pair nearer to each other (i.e., having similar hidden vector representations), and to map each negatively connected node pair more far apart from each other (i.e., having rather different hidden vector representations), in the low-dimensional embedding space. Thus, the important structural balance property of the signed networks can be well captured by the embedding representations. Then, we apply vector-based machine learning algorithms on the node vector representations learned by DNE-SBP, to carry out two important signed network mining tasks, namely, link sign prediction [29]–[31] and community detection [13], [32]–[34]. The contributions of this paper can be summarized as follows.

1) We propose a novel DNE-SBP model for signed network embedding, which leverages a semisupervised SAE to learn low-dimensional nonlinear graph representations.
2) By reconstructing the adjacency matrix, the learned hidden vector representations can capture the positive, negative, and unobserved network connections in the original network.
3) By designing the pairwise constraints to map the positively connected nodes nearer than the negatively connected nodes, the structural balance property of the signed networks can be well preserved by the embedding vector representations.
4) To deal with the highly imbalanced data in the real-world signed networks, we impose larger penalty and stronger pairwise constraint on the negative links to make them have very distinctive embedding vector representations with respect to the positive links.
5) Extensive experiments on real-world datasets demonstrate the superiority of the proposed DNE-SBP model over the state-of-the-art network embedding algorithms for graph representation learning in the signed networks.

The rest of this paper is organized as follows. Section II reviews the state-of-the-art network embedding algorithms and the semisupervised learning techniques. Section III introduces the detailed framework of DNE-SBP. Section IV reports the experimental results of DNE-SBP for link sign prediction and community detection in three real-world signed network datasets. Section V concludes this paper.

## II. Related Work

In this section, we first review the state-of-the-art network embedding algorithms developed for unsigned networks and signed networks. Then, we briefly introduce several related semisupervised learning techniques.

### A. Network Embedding Algorithms for Unsigned Networks

First, we introduce the state-of-the-art network embedding algorithms designed for unsigned networks. Perozzi *et al.* [1] proposed a DeepWalk algorithm to leverage truncated random walks to transform a graph structure into a collection of node sequences. Then, the Skip-Gram language model [35] in NLP was extended to learn the latent representations for the nodes, by considering each node in a network as a word in a document. Tang *et al.* [2] developed a Line algorithm to learn the low-dimensional vertex representations, which can preserve the first-order and the second-order network proximities. In DeepWalk and Line, a depth-first sampling (DFS) and a breadth-first sampling (BFS) strategy were adopted, respectively, to define the rigid notions of neighborhoods. In order to define the flexible notions of neighborhoods, Grover and Leskovec [3] proposed a node2vec algorithm, which adopts a flexible neighborhood sampling strategy to smoothly interpolate between DFS and BFS. Besides the "shallow" network embedding models, most recently, some promising deep network embedding algorithms have also been proposed. For example, Tian *et al.* [20] employed a sparse SAE to learn deep network representations for network clustering. Cao *et al.* [5] introduced a DNGR model which utilizes a denoising SAE to learn nonlinear network representations with high-order proximities preservation. Wang *et al.* [4] proposed an SDNE model to employ a semisupervised SAE to map the directly connected nodes near to each other in the embedding space. This SDNE model only preserves the first-order and the second-order network proximities. To capture high-order network proximities, Shen and Chung [6] developed a DNE-APP model to employ a semisupervised SAE to make the node pairs possessing higher aggregated proximities have more similar hidden vector representations. Yang *et al.* [25] adopted a semisupervised SAE to learn network representations for community detection, by mapping the nodes belonging to the same community near to each other in the embedding space. By viewing samples in a domain as nodes in a graph, Li *et al.* [36] proposed a domain adaptation framework to preserve structure consistency by learning similar feature representations for the samples belonging to the same class.

The unsigned network embedding algorithms only preserve similarities between nodes in a network. However, in the signed networks, the negative links capturing dissimilarities have extremely distinct properties with respect to the positive links [8], [9]. Thus, existing unsigned network embedding algorithms cannot be directly applied to the signed networks.

### B. Network Embedding Algorithms for Signed Networks

Next, we review several spectral embedding algorithms designed for the signed networks. Kunegis *et al.* [7] introduced a signed Laplacian matrix by extending the conventional Laplacian matrix [37] designed for unsigned networks. Then, a signed network can be embedded into a *d*-dimensional space spanned by the top-*d* eigenvectors corresponding to the smallest eigenvalues of the signed Laplacian matrix. Chiang *et al.* [13] proposed a multilevel clustering framework based on the balanced normalized cut objective, which is proved to be mathematically equivalent to the weighed kernel *k*-means clustering objective. However, this clustering framework only outputs the partitions of a network rather than an embedded map. Zheng and Skillicorn [17] proposed a spectral embedding algorithm for the signed networks. They first defined the simple normalized signed (SNS) graph Laplacian matrix and the balanced normalized signed (BNS) graph Laplacian matrix. Then, the top-*d* eigenvectors corresponding to the smallest nonzero eigenvalues of the SNS and BNS Laplacian matrices, respectively, were employed to construct a *d*-dimensional embedding space. Hsieh *et al.* [19] proposed to utilize singular value projection to complete the adjacency matrix of a given signed network. Then, the top-*d* eigenvectors of the completed adjacency matrix were employed as the low rank embeddings. However, a recent study [18] has shown that the eigenvector encoding of the cluster structure does not necessarily correspond to the smallest eigenvalues. Thus, the standard spectral clustering techniques based on the top-*k* eigenvectors associated with the smallest eigenvalues might fail to guarantee the recovery of the ground truth cluster structures. To address this issue, Mercado *et al.* [18] proposed to use the geometric mean of the Laplacian matrices, instead of the arithmetic mean used by the standard spectral clustering methods. However, measuring the geometric mean of the Laplacian matrix is computationally expensive, thus limiting this method to be scaled to the large sparse networks. In addition, Li *et al.* [38] proposed a low-rank discriminant embedding model for multiview learning which can preserve both global discriminant information and local manifold structure. A low-rank constraint was designed to map the (similar) samples belonging to the same class closer while mapping the (dissimilar) samples from different classes more separable in the low-rank space, the idea is similar to signed network embedding.

Recently, Wang *et al.* [39] proposed an SiNE algorithm to utilize a deep learning framework to learn network embeddings for the signed networks, based on the extended structural balance theory [16]. First, for each node $v_i$, a set of triplets $\{(v_i, v_j, v_k) | e_{ij} = 1, e_{ik} = -1\}$ were randomly sampled from a given signed network, where $v_j$ and $v_k$ denote a positive neighbor and a negative neighbor of $v_i$, respectively. Then, based on the sampled triplets, the goal of SiNE is to make the similarity between the hidden vector representations of a node and its positive neighbor larger than that between the node and its negative neighbor. Since SiNE learns network representations based on the sampled triplets rather than the whole network connections, some important information in the original network might be easily missing. For example, for the nodes with a very large degree, sampling a limited number of triplets might fail to get enough information to learn informative feature representations. In addition, such sampled triplets only capture observed connections, while ignoring all the unobserved connections. Thus, the network representations learned by SiNE would fail to distinguish the disconnected nodes from the connected ones. In contrast to SiNE, the proposed DNE-SBP model learns network representations from the adjacency matrix, which captures not only the positive and negative connections but also the unobserved connections. Thus, the network representations learned by DNE-SBP can not only distinguish the positively connected nodes from the negatively connected nodes but also differentiate the connected nodes from the disconnected ones. Another related work to our proposed model is the state-of-the-art deep network embedding (SDNE) model [4], which employs a semisupervised SAE to map the connected node pairs near to each other in the embedding space, without differentiating the negative connections from the positive ones. Thus, SDNE fails to capture the important structural balance property [14]–[16] of the signed networks, which suggests that the positively connected nodes should sit closer (i.e., possessing higher proximity) than the negatively connected nodes. To well capture such property, in the proposed DNE-SBP model, we carefully designed the pairwise constraints to map the positively connected nodes much nearer than the negatively connected nodes. In addition, to deal with the highly imbalanced data in the real-world signed networks, we imposed larger penalty on the reconstruction errors of the scarce negative links. Thus, the embedding representations learned by DNE-SBP would be more distinguishable between the negative links and positive links, which are indispensable information in signed networks. To the best of our knowledge, the proposed DNE-SBP model is the first work to take advantage of the semisupervised SAE for signed network embedding.

### C. Semisupervised Learning

In the real-world applications, acquiring the fully labeled data is generally very expensive and time-consuming while it is much easier to obtain the unlabeled data. Semisupervised learning is an effective technique to leverage both the limited labeled data and the abundant unlabeled data to improve the learning performance. The semisupervised learning techniques can be grouped into two categories, i.e., semisupervised classification and semisupervised clustering. On one hand, semisupervised classification [40], [41] explores how to utilize a large amount of unlabeled samples as the extra training data to improve the classification performance, such as self-training and co-training [42]–[45]. On the other hand, semisupervised clustering [46], [47] studies how to incorporate prior information, such as pairwise constraints, to boost the clustering performance. For example, Klein *et al.* [27] proposed a semisupervised clustering algorithm by incorporating the must-link (ML) and cannot-link (CL) pairwise constraints into the clustering process. The ML pairwise constraint indicates that the two instances are similar and should belong to the same cluster. While the CL pairwise constraint reflects that the two instances are dissimilar and cannot be assigned to the same cluster. Yu *et al.* [28] proposed a transitive closure-based constraint propagation approach to fully utilize the ML and CL pairwise constraints in an ensemble framework for semisupervised clustering. He *et al.* [26] developed a semisupervised clustering algorithm to propagate the ML and CL pairwise constraints through multilevel random walks. In addition, in the recent deep network embedding models [4], [6], [25], pairwise constraints have been incorporated into the SAEs to capture the proximity between nodes, which are similar to the ML constraints in semisupervised clustering. However, none of the existing deep network embedding models have utilized the CL pairwise constraints to capture the dissimilarity between the nodes. To well capture the structural balance property of the signed networks, the ML and CL pairwise constraints have been incorporated into the proposed DNE-SBP model to target for the positively and negatively connected nodes, respectively.

## III. Deep Network Embedding Model With Structural Balance Preservation

In this section, we introduce how an SAE is employed to reconstruct the adjacency matrix, how the pairwise constraints are designed, and how the DNE-SBP model can be optimized. For clarity, we summarize the frequently used notations and corresponding descriptions in Table I.

Given a signed network $G = (V, E)$ with a set of nodes $V = \{v_i\}_{i=1}^n$ and a set of edges $E = \{e_{ij}\}$, the associated signed adjacency matrix $A \in R^{n \times n}$ is defined as

$$A_{ij} = \begin{cases} 1, & \text{if relation of } (v_i, v_j) \text{ is positive} \\ -1, & \text{if relation of } (v_i, v_j) \text{ is negative} \\ 0, & \text{if relation of } (v_i, v_j) \text{ is unknown.} \end{cases}$$

Then, the signed adjacency matrix $A$ can be broken into a positive part $A^+ \in R^{n \times n}$ and a negative part $A^- \in R^{n \times n}$ as: $A_{ij}^+ = \max(A_{ij}, 0)$, $A_{ij}^- = -\min(A_{ij}, 0)$, where $A_{ij}^+, A_{ij}^- \geq 0$ represent the absolute weight of positive link and negative link, respectively.

### A. Stacked Auto-Encoder

Next, we employ an SAE to reconstruct the signed adjacency matrix $A$ so as to learn the nonlinear hidden vector representations for all the nodes in the signed network $G$. An SAE consists of $l$ layers of basic auto-encoder is constructed as follows:

TABLE I
Frequently Used Notations and Descriptions

| Notations | Descriptions |
|---|---|
| $n$ | Number of nodes in the network |
| $A$ | Signed adjacency matrix of the network |
| $v_i$ | The $i$-th node in the network |
| $l$ | Number of layers in SAE |
| $X^{(k)}, \hat{X}^{(k)}$ | Input and reconstructed matrices of $k$-th layer of SAE |
| $W_1^{(k)}, W_2^{(k)}$ | Encoding and decoding weight matrices of $k$-th layer of SAE |
| $B_1^{(k)}, B_2^{(k)}$ | Encoding and decoding bias matrices of $k$-th layer of SAE |
| $H^{(k)}$ | Hidden matrix representation learned by $k$-th layer of SAE |
| $H_i^{(k)}$ | Hidden vector representation of $v_i$ learned by $k$-th layer of SAE |
| $d(k)$ | Dimensionality of $k$-th hidden layer of SAE |
| $\gamma$ | Ratio of reconstruction penalty and pairwise constraint on negative links over positive links |

$$H^{(k)} = f\left(X^{(k)}\left(W_1^{(k)}\right)^T + B_1^{(k)}\right), k = 1, \ldots, l \quad (1)$$

$$\hat{X}^{(k)} = f\left(\hat{H}^{(k)}\left(W_2^{(k)}\right)^T + B_2^{(k)}\right), k = l, \ldots, 1 \quad (2)$$

where (1) and (2) represent the encoding and decoding process at the $k$th layer of SAE, respectively. $H^{(k)} \in R^{n \times d(k)}$ denotes the hidden matrix representation learned by the $k$th layer of SAE, $n$ is the number of nodes in network $G$, and $d(k)$ represents the dimensionality of the $k$th hidden layer of SAE. Specifically, the $i$th row of $H^{(k)}$, i.e., $H_i^{(k)} \in R^{1 \times d(k)}$ represents the hidden vector representation of node $v_i$, learned by the $k$th layer of SAE. $X^{(k)} \in R^{n \times d(k-1)}$ denotes the input matrix of the $k$th layer of SAE, $X^{(1)} = A$ and $X^{(k)} = H^{(k-1)}, \forall k = 2, \ldots, l$ indicating that the hidden matrix representation learned by the $(k-1)$th layer of SAE are utilized as the input to the $k$th layer of SAE. $W_1^{(k)} \in R^{d(k) \times d(k-1)}$ and $B_1^{(k)} \in R^{n \times d(k)}$ refer to the encoding weight and encoding bias matrices associated with the $k$th layer of SAE, respectively. In addition, $\hat{X}^{(k)} \in R^{n \times d(k-1)}$ and $\hat{H}^{(k)} \in R^{n \times d(k)}$ indicate the reconstructed matrices of $X^{(k)}$ and $H^{(k)}$, respectively, where $\hat{H}^{(l)} = H^{(l)}$ and $\hat{H}^{(k)} = \hat{X}^{(k+1)}, \forall k = l-1, \ldots, 1$. $W_2^{(k)} \in R^{d(k-1) \times d(k)}$ and $B_2^{(k)} \in R^{n \times d(k-1)}$ denote the decoding weight and the decoding bias matrices associated with the $k$th layer of SAE, respectively. $f$ is a nonlinear activation function, in the proposed DNE-SBP model, the *tanh* function $f(x) = [(e^x - e^{-x})/(e^x + e^{-x})]$ is employed as the activation function for each layer of SAE.

Then, by minimizing the reconstruction errors $\|\hat{A} - A\|_F^2$, we can learn the low-dimensional hidden vector representations which can best preserve the original network connections. However, the connections in the real-world networks are generally rather sparse, thus there are much more zero elements than nonzero elements in the adjacency matrix $A$. In such a case, directly reconstructing matrix $A$ would make the auto-encoder tend to reconstruct the zero elements (i.e., unknown connections) more than nonzero elements (i.e., observed connections). However, reconstructing observed connections should be more meaningful than reconstructing unobserved ones. To address the sparsity issue, we follow the approach in [4] to add a larger penalty on the reconstruction errors of nonzero elements. Moreover, unlike SDNE [4] which lacks the consideration about negative links, the proposed DNE-SBP model targets for the signed networks. A recent study has shown that forming negative links requires higher cost than forming positive links [12], which leads to overwhelmingly positive links in the real-world signed networks. To handle such highly imbalanced data condition in the signed networks, we design the following penalty matrix $P \in R^{n \times n}$ to make the auto-encoder focus more on reconstructing the scarce negative links than the abundant positive links

$$P_{ij} = \begin{cases} 1, & A_{ij} = 0 \\ \beta, & A_{ij} > 0 \\ \gamma * \beta, & A_{ij} < 0 \end{cases}$$

where $\beta \geq 1$ denotes the ratio of the penalty on the reconstruction errors of observed connections (i.e., nonzero elements) over that of unobserved connections (i.e., zero elements) in the input adjacency matrix $A$; $\gamma \geq 1$ indicates the ratio of penalty on the reconstruction errors of negative links over that of positive links in the input adjacency matrix $A$. By incorporating the penalty matrix $P$, we have the modified reconstruction error for the first layer of SAE as: $\mathcal{J}_1^{(1)} = (1/2n)\|(\hat{A} - A) \odot P\|_F^2$, where $\odot$ indicates an element-wise Hadamard product operator. It is worth noting that when $k \geq 2$, the input matrix of the $k$th layer of SAE is the dense hidden matrix representation learned by the $(k-1)$th layer of SAE, thus we do not incorporate the penalty matrix to address the sparsity issue for deep layers of SAE. The reconstruction errors of each $k$th layer of SAE are defined as follows:

$$\mathcal{J}_1^{(k)} = \frac{1}{2n} \begin{cases} \left\|(\hat{A} - A) \odot P\right\|_F^2, & k = 1 \\ \left\|\hat{X}^{(k)} - X^{(k)}\right\|_F^2, & k \geq 2. \end{cases} \quad (3)$$

### B. Pairwise Constraints

Next, we design a semisupervised SAE by incorporating the ML and CL pairwise constraints to capture the extended structural balance property of the signed networks. For each $k$th layer of SAE, the ML pairwise constraint $\mathcal{J}_2^{(k)}$ and the CL pairwise constraint $\mathcal{J}_3^{(k)}$ are devised as follows:

$$\mathcal{J}_2^{(k)} = \frac{1}{2n} \sum_{i=1}^{n} \sum_{j=1}^{n} A_{ij}^+ \left\|H_i^{(k)} - H_j^{(k)}\right\|_2^2 \quad (4)$$

$$\mathcal{J}_3^{(k)} = -\frac{1}{2n} \sum_{i=1}^{n} \sum_{j=1}^{n} A_{ij}^- \left\|H_i^{(k)} - H_j^{(k)}\right\|_2^2 \quad (5)$$

where $H_i^{(k)}$ and $H_j^{(k)} \in R^{1 \times d(k)}$ indicate the hidden vector representation of node $v_i$ and $v_j$, respectively, learned by the $k$th layer of SAE. On one hand, minimizing the ML pairwise constraint $\mathcal{J}_2^{(k)}$ which is equivalent to minimizing the positive ratio cut objective in signed network spectral clustering [13], we can push the positively connected nodes close to each other in the embedding space. On the other hand, via minimizing the CL pairwise constraint $\mathcal{J}_3^{(k)}$ which is equivalent to maximizing the negative ratio cut objective [13], we can pull the negatively connected nodes far away from each other in the embedding space. Moreover, to handle the highly imbalanced data (i.e., overwhelming positive links) of signed networks, we also utilize the parameter $\gamma$ to integrate $\mathcal{J}_2^{(k)}$ and $\mathcal{J}_3^{(k)}$, as follows:

$$\begin{aligned}\mathcal{J}_2^{(k)} + \gamma \mathcal{J}_3^{(k)} &= \frac{1}{n}\left(\mathrm{Tr}\left(\left(H^{(k)}\right)^T L^+ H^{(k)}\right)\right. \\ &\left. - \gamma \mathrm{Tr}\left(\left(H^{(k)}\right)^T L^- H^{(k)}\right)\right) \\ &= \frac{1}{n}\mathrm{Tr}\left(\left(H^{(k)}\right)^T L H^{(k)}\right) \end{aligned} \quad (6)$$

where $\mathrm{Tr}(.)$ denotes the trace of a matrix; $L^+ \in R^{n \times n}$ is the Laplacian matrix of $A^+$ and $L^+ = D^+ - A^+$, $D^+ \in R^{n \times n}$ denotes the diagonal degree matrix of $A^+$, with the diagonal entries $D_{ii}^+ = \sum_{j=1}^{n} A_{ij}^+$ representing the positive degree of node $v_i$. Similarly, $L^- = D^- - A^-$ is the Laplacian matrix of $A^-$, and $D^-$ indicates the diagonal degree matrix of $A^-$, with $D_{ii}^- = \sum_{j=1}^{n} A_{ij}^-$ denoting the negative degree of node $v_i$. In addition, $L = L^+ - \gamma L^-$, where $\gamma \geq 1$ indicates the ratio of weight of the CL pairwise constraint targeting for the negative links over that of the ML pairwise constraint targeting for the positive links. A larger value of $\gamma$ would make DNE-SBP tend to enlarge the distance between the negatively connected nodes more, with respect to narrowing the distance between the positively connected nodes. Note that when $\gamma = 1$ minimizing $(\mathcal{J}_2^{(k)} + \mathcal{J}_3^{(k)})$ is analogous to minimizing the Rayleigh quotient of SNS graph Laplacian in signed network spectral embedding [17]. This reflects that our designed pairwise constraints are indeed able to capture and preserve the extended structural balance property of the signed networks. In addition, note that in the pairwise constraint (6) and the penalty-modified reconstruction errors (3), we set the same value to the parameter $\gamma$ to make the DNE-SBP model easily distinguish the scarce negative links from the abundant positive links.

### C. Overall Loss Function

By integrating the reconstruction errors (3), the pairwise constraint (6), and a $L2$-norm regularization term $\mathcal{J}_4^{(k)} = (1/2)(\|W_1^{(k)}\|_F^2 + \|W_2^{(k)}\|_F^2)$, the overall loss function of DNE-SBP is defined as

$$\begin{aligned}\mathcal{J} &= \sum_{k=1}^{l} \mathcal{J}^{(k)} \\ &= \sum_{k=1}^{l}\left(\mathcal{J}_1^{(k)} + \alpha_k\left(\mathcal{J}_2^{(k)} + \gamma_k \mathcal{J}_3^{(k)}\right) + \lambda_k \mathcal{J}_4^{(k)}\right) \quad (7)\end{aligned}$$

where $l$ represents the number of layers in the SAE; $\mathcal{J}^{(k)}$ denotes the loss function of the $k$th layer of SAE; $\alpha_k, \lambda_k > 0$ refer to the weights of the pairwise constraint and the regularization term at the $k$th layer of SAE, respectively. By minimizing (7), we can learn the hidden node vector representations which not only preserve the original network connections but also capture the structural balance property of the signed network.

### D. Optimization of DNE-SBP

Here, we explain how to optimize DNE-SBP. First, for each $k$th layer basic auto-encoder, with the help of back-propagation algorithm, the "error" terms of its output layer $\delta_3^{(k)} \in R^{n \times d(k-1)}$ and hidden layer $\delta_2^{(k)} \in R^{n \times d(k)}$ can be computed, respectively, as

$$\delta_3^{(k)} = \begin{cases} (\hat{A} - A) \odot P \odot P \odot f'(Z_3^{(k)}), & k = 1 \\ (\hat{X}^{(k)} - X^{(k)}) \odot f'(Z_3^{(k)}), & k \geq 2 \end{cases} \quad (8)$$

$$\delta_2^{(k)} = \left(\delta_3^{(k)} W_2^{(k)} + \alpha_k (L + L^T) H^{(k)}\right) \odot f'(Z_2^{(k)}) \quad (9)$$

where $Z_3^{(k)} = H^{(k)}(W_2^{(k)})^T + b_2^{(k)}$, $Z_2^{(k)} = X^{(k)}(W_1^{(k)})^T + b_1^{(k)}$, $H^{(k)} = f(Z_2^{(k)})$, $\hat{X}^{(k)} = f(Z_3^{(k)})$; and $f'$ denotes the derivative of the activation function.

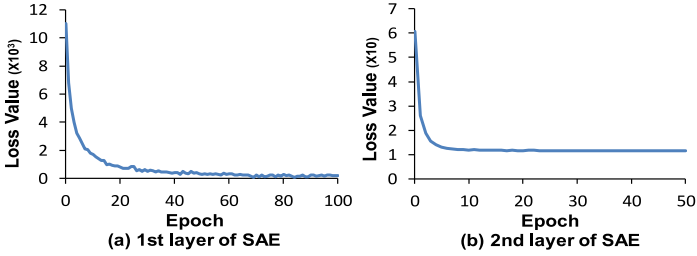

Fig. 1. Convergence of the first layer and the second layer of SAE in the DNE-SBP model in the Wiki dataset.

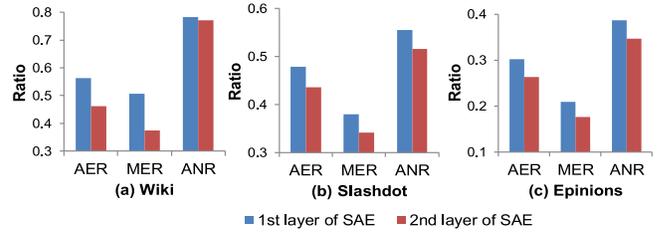

Fig. 2. AER, MER, and ANR ratios of the distances between the positively connected nodes over that between the negatively connected nodes in the embedding spaces learned by the first layer and the second layer of SAE in the DNE-SBP model.

Next, the partial derivative with respect to the encoding weight $W_1^{(k)}$, decoding weight $W_2^{(k)}$, encoding bias $B_1^{(k)}$, and decoding bias $B_2^{(k)}$, can be computed, respectively, as follows:

$$\frac{\partial \mathcal{J}^{(k)}}{\partial W_1^{(k)}} = \frac{1}{n} \left(\delta_2^{(k)}\right)^T X^{(k)} + \lambda_k W_1^{(k)}$$

$$\frac{\partial \mathcal{J}^{(k)}}{\partial W_2^{(k)}} = \frac{1}{n} \left(\delta_3^{(k)}\right)^T H^{(k)} + \lambda_k W_2^{(k)}$$

$$\frac{\partial \mathcal{J}^{(k)}}{\partial B_1^{(k)}} = \delta_2^{(k)}/n, \quad \frac{\partial \mathcal{J}^{(k)}}{\partial B_2^{(k)}} = \delta_3^{(k)}/n. \quad (10)$$

To minimize the loss function $\mathcal{J}^{(k)}$, we use stochastic gradient descent to update the parameters as: $W_1^{(k)} = W_1^{(k)} - \eta_k[(\partial \mathcal{J}^{(k)})/(\partial W_1^{(k)})]$; $W_2^{(k)} = W_2^{(k)} - \eta_k[(\partial \mathcal{J}^{(k)})/(\partial W_2^{(k)})]$; $B_1^{(k)} = B_1^{(k)} - \eta_k[(\partial \mathcal{J}^{(k)})/(\partial B_1^{(k)})]$; $B_2^{(k)} = B_2^{(k)} - \eta_k[(\partial \mathcal{J}^{(k)})/(\partial B_2^{(k)})]$, where $\eta_k$ indicates the learning rate of the $k$th layer basic auto-encoder. Next, to optimize an SAE consists of multiple layers of basic auto-encoder, we adopt a greedy layer-wise training approach, as in [5], [20], and [48]. The time complexity of DNE-SBP is $O(nchI)$, where $n$ denotes the number of nodes in the network; $c$ indicates the average degree of the network; $h$ represents the maximum dimensionality of the hidden layers in SAE, i.e., $h = d(1)$; $I$ refers to the number of iterations. Since $chI$ is independent of $n$, the overall time complexity of DNE-SBP is linear to the number of nodes in the network.

## IV. EXPERIMENTS

### A. Datasets

We evaluated the graph representation learning performance of the proposed DNE-SBP model for link sign prediction and community detection in three real-world signed networks, namely Epinions, Slashdot, and Wiki. The Epinions dataset [49] is a "who trust whom" online social network generated from the Epinions site, where one user can "trust" (positive) or "distrust" (negative) another. The Slashdot dataset [50] is a signed social network extracted from the technology news site Slashdot, where users can form the relationships as friends (positive) or foes (negative). The Wiki dataset [51] is extracted from the Wikipedia site, which describes the votes "for" (positive) and "against" (negative) the other in elections. In the experiments, we used the full Wiki dataset, and extracted 7000 nodes with the largest degree and retained all the edges between the selected nodes from the original Epinions and Slashdot datasets. Table II shows the statistics of the three datasets. Among three networks, Wiki is the sparsest one, Slashdot is second sparsest, while Epinions is the densest.

TABLE II
STATISTICS OF SIGNED NETWORK DATASETS

| Datasets | Epinions | Slashdot | Wiki |
|---|---|---|---|
| # Users | 7000 | 7000 | 7118 |
| # Links | 451149 | 238029 | 103675 |
| # Positive links | 404006 | 181354 | 81318 |
| # Negative links | 47143 | 56675 | 22357 |

TABLE III
LAYER CONFIGURATION OF THE SAE ON THREE DATASETS FOR LINK SIGN PREDICTION AND COMMUNITY DETECTION

| Datasets | Dimensionality of each layer of SAE | |
|---|---|---|
| | Link Sign Prediction | Community Detection |
| Wiki | 7118-256-64 | 7118-512-256-128-64 |
| Slashdot | 7000-256-64 | 7000-512-256-128-64 |
| Epinions | 7000-256-64 | 7000-512-256-128-64 |

### B. Implementation Details

For link sign prediction, we built a two-layer SAE in the DNE-SBP model. The layer configurations of the SAE for the three datasets are shown in Table III. For example, for Wiki dataset, the layer configuration "7118-256-64" indicates that the first layer basic auto-encoder and the second layer basic auto-encoder in the SAE were configured with the dimensionality of each layer as 7118-256-7118 and 256-64-256, respectively. In addition, the batch sizes of the first layer and the deeper layers of SAE were set as 500 and 100, respectively, and the learning rates were set as $\eta_1 = 0.025$ and $\eta_k = 0.015$, $\forall k \geq 2$. We set the weight of $L2$-norm regularization as $\lambda_1 = 0.05$ and $\lambda_k = 0.1$, $\forall k \geq 2$ in both Epinions and Slashdot datasets; and set $\lambda_1 = 0.05$ and $\lambda_k = 0.25$, $\forall k \geq 2$ in the Wiki dataset. In addition, we set the weight of pairwise constraint at the first layer and the deeper layers of SAE as: $\alpha_1 = 16, 14, 10$ and $\alpha_k = 0.4, 0.2, 0.2, \forall k \geq 2$ for the Wiki, Slashdot, and Epinions datasets, respectively. The ratio of penalty on the reconstructing errors of nonzero input elements over that of zero input elements were set as $\beta = 25, 25$ and 10 in the Wiki, Slashdot, and Epinions datasets, respectively. Besides, we set the ratio of penalty for the reconstructing errors of negative links over that of positive links, and the ratio of the CL constraint weight over that of ML constraint weight as $\gamma_1 = \text{floor}([(\sum_{i=1}^n \sum_{j=1}^n A_{ij}^+)/(\sum_{i=1}^n \sum_{j=1}^n A_{ij}^-)])$, where floor($x$) rounds $x$ to the nearest integer less than or equal to $x$. In addition, note that we set $\gamma_k = 1, \forall k \geq 2$, because after the first layer embedding, the larger penalty and stronger pairwise constraint for negative links have already been imposed to learn the hidden representations.

On the other hand, for community detection, a four-layer SAE was built for the three datasets, as shown in Table III. The parameter settings for community detection are similar to those introduced for link sign prediction. Here, we only report the differences. First, in contrast to link sign prediction, we set a larger batch size of 1000 for each layer of SAE. Second, we assigned a larger weight to the pairwise constraint at the deeper layers of SAE, i.e., $\alpha_k = 1.5, \forall k \geq 2$. This is because the structural balance theory [14], [15] was originally proposed for network clustering, thus it should be more important and necessary to preserve the structural balance property for the community detection task (by assigning larger weight to the pairwise constraint). In addition, for both link sign prediction and community detection, we employed the deepest hidden vector representations learned by the last layer of SAE as the node vector representations, with the dimensionality $d = 64$.

Each layer of SAE was optimized using stochastic gradient descent with the learning rates introduced above. As shown in Fig. 1(a), the value of the loss function of the first layer of SAE decreases rapidly at the first 20 iterations and then gradually converges at 80 iterations. While for the second layer of SAE, as shown in Fig. 1(b), the loss function is easier to get convergence after less than 20 iterations.

### C. Analysis of Embedding Learned by DNE-SBP

Here, we analyze whether the network representations learned by the proposed DNE-SBP model can preserve the extended structural balance property of the signed networks, i.e., whether the positively connected nodes are sitting closer than the negatively connected nodes in the embedding space. In this regard, we adopted three distance measures introduced in [17], namely average edge ratio (AER), median edge ratio (MER), and average node ratio (ANR). AER is defined as the ratio of the average embedded distance between the positively connected nodes over that between the negatively connected nodes, i.e.,

$$\frac{\left(\sum_{i=1}^n \sum_{j=1}^n A_{ij}^+ d_{ij}\right)/\sum_{i=1}^n \sum_{j=1}^n A_{ij}^+}{\left(\sum_{i=1}^n \sum_{j=1}^n A_{ij}^- d_{ij}\right)/\sum_{i=1}^n \sum_{j=1}^n A_{ij}^-}$$

where $d_{ij}$ indicates the Euclidean distance between the hidden vector representations of node $v_i$ and $v_j$. MER is the ratio of the median of the embedded distance between the positively connected nodes over that between the negatively connected nodes, defined as follows:

$$\frac{\text{median}\left(d_{ij}|A_{ij}^+ > 0\right)}{\text{median}\left(d_{ij}|A_{ij}^- > 0\right)}.$$

ANR is the ratio of the average embedded length of positive links over that of negative links, from the perspective of nodes, defined as follows:

$$\frac{\sum_{i=1}^n \left(\sum_{j=1}^n A_{ij}^+ d_{ij}/D_{ii}^+\right)/n_p}{\sum_{i=1}^n \left(\sum_{j=1}^n A_{ij}^- d_{ij}/D_{ii}^-\right)/n_n}$$

where $n_p$ and $n_n$ indicate the number of nodes having at least one positive link and having at least one negative link, respectively. As shown in Fig. 2, all the three ratios are smaller than 1 in three datasets, indicating that the embeddings learned by DNE-SBP can actually capture the extended structural balance property. Moreover, we can observe that the ratios measured in the embedding space learned by the second layer of SAE were smaller than that learned by the first layer of SAE. This indicates that the hidden representations learned by the deeper layer of SAE can better satisfy the extended structural balance condition.

### D. Baseline Algorithms

The following state-of-the-art network embedding algorithms were employed to benchmark against the proposed DNE-SBP model.
1) *SL [7]:* It is a spectral clustering algorithm, with a signed Laplacian matrix defined as $\bar{L} = \bar{D} - A$, where $\bar{D}_{ii} = \sum_{j=1}^n |A_{ij}|$ indicates the sum of positive and negative degree of node $v_i$. The top-$d$ eigenvectors of $\bar{L}$ are selected as the node vector representations.
2) *SNS [17]:* It is a spectral embedding algorithm defining the SNS Laplacian matrix as $L_{\text{SNS}} = \bar{D}^{-1}(D^+ - D^- - A)$. The top-$d$ eigenvectors of $L_{\text{SNS}}$ are selected as the node vector representations.
3) *BNS [17]:* It is a spectral embedding algorithm with the BNS Laplacian matrix defined as $L_{\text{BNS}} = \bar{D}^{-1}(D^+ - A)$. The top-$d$ eigenvectors of $L_{\text{BNS}}$ are selected as the node vector representations.
4) *SiNE [39]:* It is a deep network embedding model designed to preserve the extended structural balance property of the signed networks. It randomly samples a set of triplets containing positive and negative neighbors of each node in the given network. Then, it employs a deep learning framework to learn the node vector representations based on the sampled triplets.
5) *SDNE [4]:* It is a deep network embedding model designed for unsigned networks. It employs a semisupervised SAE to map the connected node pairs close to each other, without differentiating positive and negative links.

Since the structural balance theory is naturally defined for undirected networks [8], we evaluated all the comparing algorithms in undirected signed networks. To guarantee the best performance of SiNE, we used the default settings in [39], i.e., building a three-layer neural network with the dimensionality of each layer and the node vector representations as $d = 20$. For other baselines in link sign prediction, we used the same dimensionality of node vector representation as in our DNE-SBP model, i.e., $d = 64$. In addition, note that the spectral embedding algorithms, i.e., SL, SNS, and BNS, assume that the top-$k$ eigenvectors encode the corresponding $k$-way clustering structure [7], [17]. Thus, for the community detection task, the dimensionality of the node vector representations learned by these spectral embedding algorithms were set as equal to the number of clusters in the given network, i.e., $d = k$.

### E. Experimental Results

In this section, we report the experimental results of the network embedding algorithms for link sign prediction and community detection in three real-world signed networks.

*1) Link Sign Prediction:* For link sign prediction, first a training fraction $f\%$ of edges were randomly sampled from the given network to construct the signed adjacency matrix, which is employed as the input to learn the low-dimensional node vector representations. Then, four types of edge feature vector representations were built based on the node vector representations learned by each network embedding algorithm, as in [3]

$$\begin{aligned}
\mathbf{L1}&: H(e_{ij}) = |H_i - H_j| \\
\mathbf{L2}&: H(e_{ij}) = |H_i - H_j|^2 \\
\mathbf{Had}&: H(e_{ij}) = H_i \odot H_j \\
\mathbf{Avg}&: H(e_{ij}) = (H_i + H_j)/2
\end{aligned}$$

where $H(e_{ij})$ indicates the feature vector representation of edge $e_{ij}$; $H_i$ and $H_j$ denote the feature vector representation of node $v_i$ and $v_j$, respectively. A logistic regression (LR) model was trained based on the edge vector representations and the observed edge labels in the training set. Next, the LR model was employed to predict the labels of edges in the testing set. As the signed network datasets are overwhelmingly positive, directly evaluating the accuracy on such highly imbalanced datasets will be misleading. Thus, following [29], [30], and [39], we adopted the "area under the ROC curve" (AUC) metric to evaluate the link sign prediction performance, which is insensitive to the imbalanced data. The higher the AUC score, the better the link sign prediction performance. In addition, we employed the average precision (AP) [52] as another metric. As compared to precision@$k$, AP is more concerned with the retrieved items ranking ahead. To better reflect the link sign prediction performance in such highly imbalanced datasets, we calculated the AP of the scarce class (i.e., negative links). The higher the AP, the better the link sign prediction performance. Table IV reports the AUC and AP scores of all the comparing algorithms with four types of edge representation in the Wiki dataset. Tables V and VI report the highest AUC and AP scores each algorithm can achieve with its best suitable edge representation, in the Slashdot and Epinions datasets, respectively. The reported AUC and AP scores for each comparing algorithm were averaged over the same five random splits.

As shown in Table IV, when the given network is rather sparse, e.g., just giving 20% of observed links in the sparest Wiki network, the SiNE algorithm with the Avg edge representation can achieve the highest AUC and AP scores among all the comparing algorithms. Except that, the proposed DNE-SBP model with the Had edge representation always achieved the highest AUC and AP scores in the three signed networks, as shown in Tables IV–VI. This could be explained by the fact that SiNE learns network representations based on the sampled triplets [39]. Then, if the given network is dense (i.e., having a large number of possible triplets), a limited number of samples would unavoidably lose some information. Thus, SiNE would perform worse as the given network is denser. In contrast to SiNE, the proposed DNE-SBP model learns network representations from the adjacency matrix of a given network. Thus, we can take advantage of the whole network connections to learn more informative network representations. In addition, we can observe that both DNE-SBP and SiNE outperform the spectral embedding algorithms in the three signed networks. This demonstrates the higher effectiveness of deep learning techniques for graph representation learning, as compared to the linear matrix decomposition methods.

Second, we discuss the performance of the spectral embedding algorithms, i.e., SL, SNS, and BNS. We can see that in the densest Epinions dataset (shown in Table VI), SL with the Had edge representation can achieve the highest AUC scores among all the spectral embedding baselines. However, conversely, in the sparsest Wiki dataset (shown in Table IV), SL always achieved the lowest AUC and AP scores, no matter what percentage of observed links were used for training. This reflects that SL is rather unsuitable for the sparse networks, despite of the fact that it can perform much better in the dense networks. In addition, as shown in Table IV, SNS with the L1 edge representation can achieve the highest AUC scores among all the spectral embedding methods in the Wiki dataset. While BNS can achieve both higher AUC and AP scores than SNS in both Slashdot and Epinions datasets, as shown in Tables V and VI. This reflects that BNS performs better in the dense networks, while showing greater challenge when dealing with the sparse networks.

Next, let us continue to evaluate the performance of the unsigned network embedding algorithm, i.e., SDNE. As shown in Tables V and VI, SDNE always achieved much lower AUC and AP scores than all the signed network embedding algorithms in Slashdot and Epinions datasets. This is because SDNE aims to map the connected node pairs closer to each other based on the social theory which suggests that the connected nodes tend to have similar preferences [10], [11]. However, this theory is not applicable for the signed networks where negative links indicate dissimilarity while the positive links indicate similarity. Thus, directly applying the unsigned network embedding algorithm to the signed networks would fail to capture the structural balance property [14]–[16] which requires the positively connected nodes to sit closer than the negatively connected ones.

Moreover, we can observe that all the network embedding algorithms can achieve the highest AUC scores in the

TABLE IV
AUC AND AP OF LINK SIGN PREDICTION IN THE WIKI DATASET. THE HIGHEST AUC AND AP SCORES AMONG ALL THE COMPARING METHODS ARE SHOWN IN BOLDFACE. * AND ** INDICATE STATISTICALLY SUPERIOR PERFORMANCE TO SINE (WITH ITS BEST SUITABLE EDGE FEATURE) AT LEVEL OF (0.05, 0.01) USING A PAIRED $t$-TEST

| Algorithms | Edge Features | AUC | | | | AP | | | |
|---|---|---|---|---|---|---|---|---|---|
| | | % of observed links used for training | | | | % of observed links used for training | | | |
| | | 20 | 40 | 60 | 80 | 20 | 40 | 60 | 80 |
| DNE-SBP | L1 | 0.7337 | 0.8184 | 0.8431 | 0.8517 | 0.4662 | 0.5987 | 0.6346 | 0.6389 |
| | L2 | 0.7161 | 0.7991 | 0.8319 | 0.8431 | 0.4604 | 0.5805 | 0.6238 | 0.6321 |
| | Had | 0.7937 | **0.8459** | **0.8626**** | **0.8681*** | 0.5210 | **0.6268*** | **0.6595**** | **0.6642**** |
| | Avg | 0.8038 | 0.8454 | 0.8562 | 0.8591 | 0.5311 | 0.6057 | 0.6267 | 0.6305 |
| SiNE | L1 | 0.6143 | 0.6700 | 0.6818 | 0.7021 | 0.3005 | 0.3390 | 0.3601 | 0.3671 |
| | L2 | 0.6775 | 0.7092 | 0.7155 | 0.7001 | 0.3677 | 0.3952 | 0.3965 | 0.3682 |
| | Had | 0.7961 | 0.8215 | 0.8053 | 0.8093 | 0.5212 | 0.5396 | 0.5047 | 0.5233 |
| | Avg | **0.8080** | 0.8399 | 0.8537 | 0.8644 | **0.5492** | 0.5863 | 0.6130 | 0.6305 |
| SL | L1 | 0.5726 | 0.5858 | 0.5900 | 0.6024 | 0.2771 | 0.2777 | 0.2795 | 0.2857 |
| | L2 | 0.5118 | 0.5046 | 0.5026 | 0.5012 | 0.2319 | 0.2205 | 0.2177 | 0.2164 |
| | Had | 0.5000 | 0.5000 | 0.5000 | 0.5000 | 0.2156 | 0.2133 | 0.2142 | 0.2141 |
| | Avg | 0.5001 | 0.5003 | 0.5038 | 0.5040 | 0.2232 | 0.2172 | 0.2162 | 0.2157 |
| SNS | L1 | 0.6747 | 0.6928 | 0.7011 | 0.6951 | 0.3500 | 0.3665 | 0.3768 | 0.3734 |
| | L2 | 0.6497 | 0.6595 | 0.6571 | 0.6703 | 0.3367 | 0.3508 | 0.3558 | 0.3597 |
| | Had | 0.5106 | 0.5028 | 0.4998 | 0.5018 | 0.2663 | 0.2529 | 0.2487 | 0.2635 |
| | Avg | 0.5461 | 0.5252 | 0.5156 | 0.5141 | 0.2784 | 0.2741 | 0.2543 | 0.2513 |
| BNS | L1 | 0.6541 | 0.6440 | 0.6505 | 0.6536 | 0.4073 | 0.3918 | 0.4044 | 0.4000 |
| | L2 | 0.5149 | 0.5078 | 0.5042 | 0.5025 | 0.2491 | 0.2323 | 0.2252 | 0.2199 |
| | Had | 0.5000 | 0.5000 | 0.5000 | 0.5000 | 0.2146 | 0.2128 | 0.2137 | 0.2135 |
| | Avg | 0.5122 | 0.5108 | 0.5037 | 0.5051 | 0.2504 | 0.2442 | 0.2328 | 0.2227 |
| SDNE | L1 | 0.6032 | 0.6232 | 0.6353 | 0.6374 | 0.3027 | 0.3475 | 0.3899 | 0.4072 |
| | L2 | 0.6030 | 0.6223 | 0.6307 | 0.6380 | 0.3121 | 0.3594 | 0.3893 | 0.4090 |
| | Had | 0.6494 | 0.6746 | 0.6825 | 0.6804 | 0.3456 | 0.3935 | 0.4223 | 0.4332 |
| | Avg | 0.6507 | 0.6744 | 0.6940 | 0.6977 | 0.3365 | 0.3756 | 0.4228 | 0.4464 |

TABLE V
AUC AND AP OF LINK SIGN PREDICTION IN THE SLASHDOT DATASET. THE HIGHEST AUC AND AP SCORES AMONG ALL THE COMPARING METHODS ARE SHOWN IN BOLDFACE. ** INDICATES STATISTICALLY SUPERIOR PERFORMANCE TO SINE AT LEVEL OF 0.01 USING A PAIRED $t$-TEST

| Algorithms | Best Edge Features | AUC | | | | AP | | | |
|---|---|---|---|---|---|---|---|---|---|
| | | % of observed links used for training | | | | % of observed links used for training | | | |
| | | 20 | 40 | 60 | 80 | 20 | 40 | 60 | 80 |
| DNE-SBP | Had | **0.8473 **** | **0.8855 **** | **0.8979**** | **0.9058**** | **0.6821 **** | **0.7571**** | **0.7808**** | **0.7957**** |
| SiNE | Avg | 0.8192 | 0.8467 | 0.8572 | 0.8623 | 0.5986 | 0.6427 | 0.6645 | 0.6713 |
| SL | Had | 0.7824 | 0.8350 | 0.8500 | 0.8528 | 0.5117 | 0.6065 | 0.6432 | 0.6363 |
| SNS | L1 | 0.7654 | 0.7726 | 0.7215 | 0.7712 | 0.5213 | 0.5484 | 0.5191 | 0.5745 |
| BNS | L1 | 0.7761 | 0.8023 | 0.8167 | 0.8292 | 0.5250 | 0.5864 | 0.6278 | 0.6544 |
| SDNE | Avg | 0.6672 | 0.7213 | 0.7429 | 0.7503 | 0.3813 | 0.4339 | 0.4627 | 0.4726 |

TABLE VI
AUC AND AP OF LINK SIGN PREDICTION IN THE EPINIONS DATASET. THE HIGHEST AUC AND AP SCORES AMONG ALL THE COMPARING METHODS ARE SHOWN IN BOLDFACE. ** INDICATES STATISTICALLY SUPERIOR PERFORMANCE TO SINE AT LEVEL OF 0.01 USING A PAIRED $t$-TEST

| Algorithms | Best Edge Features | AUC | | | | AP | | | |
|---|---|---|---|---|---|---|---|---|---|
| | | % of observed links used for training | | | | % of observed links used for training | | | |
| | | 20 | 40 | 60 | 80 | 20 | 40 | 60 | 80 |
| DNE-SBP | Had | **0.9137**** | **0.9288**** | **0.9336**** | **0.9373**** | **0.7280**** | **0.7686**** | **0.7846**** | **0.7925**** |
| SiNE | Avg | 0.9009 | 0.9127 | 0.9135 | 0.9114 | 0.6201 | 0.6432 | 0.6481 | 0.6491 |
| SL | Had | 0.8811 | 0.8991 | 0.9076 | 0.9031 | 0.5911 | 0.6618 | 0.6932 | 0.6842 |
| SNS | L1 | 0.8381 | 0.7092 | 0.6792 | 0.8074 | 0.5417 | 0.3524 | 0.3905 | 0.5221 |
| BNS | L2 | 0.8494 | 0.8725 | 0.8817 | 0.8908 | 0.5170 | 0.5889 | 0.6311 | 0.6617 |
| SDNE | Avg | 0.7248 | 0.7515 | 0.7635 | 0.7620 | 0.2622 | 0.2917 | 0.3101 | 0.3033 |

Epinions dataset, while the lowest AUC scores in the Wiki dataset. These could be explained by the previous findings in [29] and [53] that the structural balance condition is most satisfied in the Epinions network, while least satisfied in the Wiki network. Thus, it should be easier to predict the link signed labels based on the structural balance property for the Epinions dataset.

In addition, all the embedding algorithms were evaluated on a machine with an Intel Core i7 4.2GHz CPU and 32GB RAM. We measured the CPU time of all the embedding algorithms for learning the feature vector representations in the Epinions dataset. The running time of SDNE and our DNE-SBP model were lowest among all the comparing algorithms, about 17 min. And the running time was 33 min for SiNE,

TABLE VII
ERROR RATES (%) OF $k$-WAY CLUSTERING IN THREE SIGNED NETWORKS. THE LOWEST ERROR RATES AMONG ALL THE COMPARING ALGORITHMS ARE SHOWN IN BOLDFACE

| Dataset | Algorithms | Number of cluster $k$ | | | | | | | | | Average |
|---|---|---|---|---|---|---|---|---|---|---|---|
| | | 2 | 3 | 4 | 5 | 6 | 7 | 8 | 9 | 10 | |
| Wiki | DNE-SBP | **15.64** | **15.66** | **15.74** | **15.71** | **15.71** | **15.71** | **15.71** | **15.70** | **15.69** | **15.70** |
| | SL | 21.94 | 21.94 | 21.94 | 21.94 | 21.94 | 21.93 | 21.93 | 21.93 | 21.93 | 21.94 |
| | SNS | 21.94 | 21.94 | 21.94 | 21.94 | 21.94 | 21.94 | 21.94 | 21.94 | 21.94 | 21.94 |
| | BNS | 21.86 | 21.80 | 21.77 | 21.76 | 21.74 | 21.72 | 21.72 | 21.71 | 21.70 | 21.75 |
| | SiNE | 33.48 | 53.57 | 58.62 | 61.14 | 62.01 | 63.30 | 66.85 | 67.74 | 69.12 | 59.54 |
| | SDNE | 26.01 | 33.67 | 42.16 | 42.69 | 44.97 | 55.81 | 56.03 | 55.70 | 55.84 | 45.88 |
| Slashdot | DNE-SBP | **18.54** | **18.19** | **18.15** | **18.17** | **18.15** | **18.23** | **18.22** | **18.23** | **18.23** | **18.24** |
| | SL | 25.38 | 25.39 | 25.39 | 25.39 | 25.39 | 25.39 | 25.39 | 25.40 | 25.40 | 25.39 |
| | SNS | 25.36 | 25.24 | 25.21 | 25.20 | 25.20 | 25.06 | 25.05 | 25.05 | 24.90 | 25.14 |
| | BNS | 25.19 | 25.15 | 25.00 | 24.81 | 24.74 | 24.84 | 24.81 | 24.44 | 23.49 | 24.72 |
| | SiNE | 40.38 | 47.65 | 51.18 | 57.00 | 55.89 | 56.16 | 58.37 | 62.35 | 63.72 | 54.74 |
| | SDNE | 48.91 | 55.75 | 57.78 | 61.29 | 62.34 | 63.65 | 64.00 | 65.35 | 66.09 | 60.57 |
| Epinions | DNE-SBP | **7.95** | **8.25** | **8.20** | **8.29** | **8.30** | **8.33** | **8.31** | **8.31** | **8.32** | **8.25** |
| | SL | 12.20 | 12.20 | 12.20 | 12.20 | 12.20 | 12.20 | 12.21 | 12.21 | 12.21 | 12.20 |
| | SNS | 11.73 | 12.08 | 12.10 | 11.93 | 11.97 | 11.78 | 11.93 | 12.05 | 11.90 | 11.94 |
| | BNS | 9.01 | 9.01 | 9.77 | 9.02 | 9.57 | 9.54 | 9.50 | 9.54 | 9.55 | 9.39 |
| | SiNE | 40.70 | 48.03 | 56.43 | 61.87 | 66.58 | 68.21 | 69.64 | 70.82 | 71.61 | 61.54 |
| | SDNE | 36.76 | 38.75 | 43.19 | 48.35 | 49.26 | 50.02 | 52.89 | 56.34 | 57.01 | 48.06 |

22.4, 30.3, and 30.8 min for the SL, SNS, and BNS spectral embedding algorithms, respectively. Thus, the proposed DNE-SBP model not only achieves the highest AUC and AP scores, but also runs efficiently.

*2) Community Detection:* For community detection, a $k$-means algorithm was run on the node vector representations learned by each network embedding algorithm to get the clustering results. Note that unlike community detection in unsigned networks, the objective of signed network clustering is to group nodes into $k$ clusters, where the connections between the nodes within the same cluster should be mostly positive, while the connections between the nodes belonging to different clusters should be mostly negative [8], [13]. Thus, to evaluate the performance of signed network community detection, we adopted the "error rate" metric, which is widely utilized in [13], [19], and [32]. The error rate $E$ is defined as the sum of the number of negative edges within the same cluster and the number of positive edges between different clusters, normalized by the total number of edges in the network, as follows:

$$E = \frac{\sum_{i=1}^{n} \sum_{j=1}^{n} A_{ij}^{-} \delta(c_i, c_j) + A_{ij}^{+} (1 - \delta(c_i, c_j))}{\sum_{i=1}^{n} \sum_{j=1}^{n} |A_{ij}|}$$

where $c_i, c_j$ indicate the community node $v_i$ and $v_j$ belonging to, respectively; if $v_i$ and $v_j$ are assigned to the same community, then $\delta(c_i, c_j) = 1$, otherwise, $\delta(c_i, c_j) = 0$. The lower the error rate, the better the signed network clustering performance.

First, as shown in Table VII, the proposed DNE-SBP model always outperformed all the baselines (i.e., achieved the lowest error rates) in all the three datasets. This again proves that the network representations learned by DNE-SBP can well capture and preserve the structural balance property of signed networks. In addition, among all the spectral embedding algorithms, BNS achieved the lowest error rates in all the three datasets. SNS achieved slightly lower error rates than SL in the Slashdot and Epinions networks. Note that the objective function of SNS is analog to the pairwise constraint designed in the proposed DNE-SBP model. However, when learning network representations, SNS employs EVD to linearly project the original network into a low-dimensional embedding space [17]. In contrast, we take advantage of deep learning technique to learn nonlinear network representations. Thus, the significant outperformance of DNE-SBP with respect to SNS reflects that the deep learning techniques possess more powerful feature representation learning ability to capture the complex underlying network structures.

Second, we can see that SiNE achieved the highest error rates among all the comparing algorithms in the Wiki and Epinions datasets. As the number of cluster $k$ increases, SiNE would perform even worse. It might be explained by the fact that SiNE learns network representations based on the triplets sampled from the observed connections, while all the unobserved connections in the original network are ignored. Thus, the network representations learned by SiNE would fail to distinguish the disconnected nodes from the connected nodes, which is not desirable for the community detection task. In addition, the sampled triplets only capture the local neighborhood structure, however, community detection also requires the global structural information. Hence, SiNE performs rather badly when learning network representations for signed network community detection.

Moreover, we can see that SDNE performed significantly worse (i.e., achieved much higher error rates) than all the spectral embedding algorithms developed for the signed networks. This again confirms that the unsigned network embedding algorithm fails to learn informative network representations for the signed networks. Thus, it is indeed necessary to design the network embedding algorithms targeting for the signed networks, which can well capture the important structural balance property so as to easily distinguish negative links from positive links.

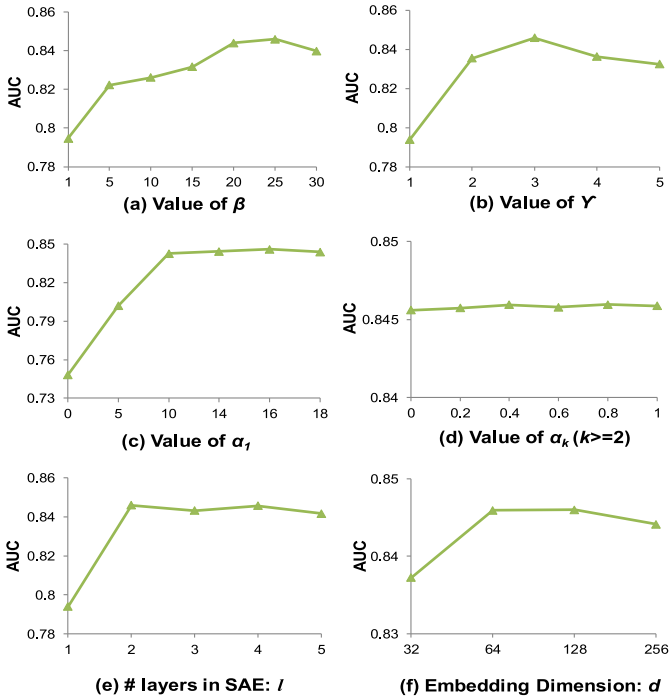

Fig. 3. Sensitivity of the parameters $\beta, \gamma, \alpha, l, d$ on the AUC score of DNE-SBP (with the Had edge representation) for link sign prediction, when 40% of observed links were used for training in the Wiki dataset. The higher the AUC score, the better the performance.

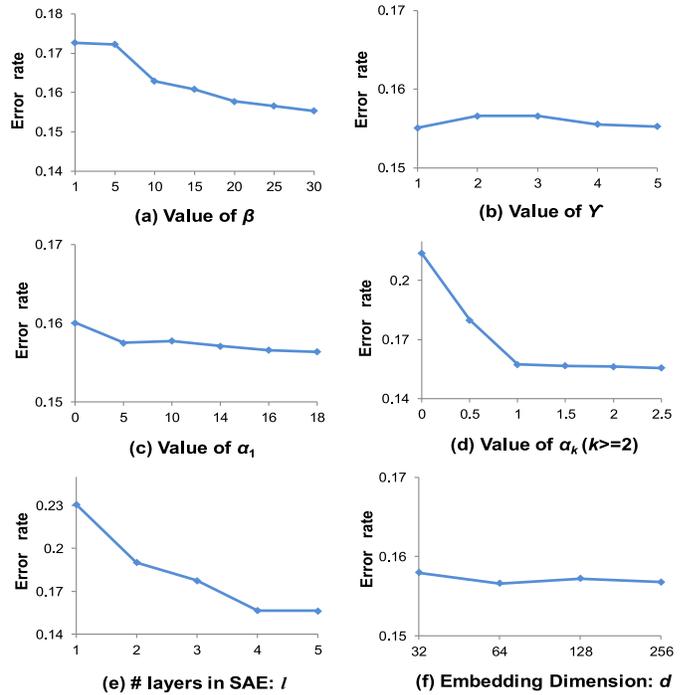

Fig. 4. Sensitivity of the parameters $\beta, \gamma, \alpha, l, d$ on the error rate of DNE-SBP for 3-way signed network community detection in the Wiki dataset. The lower the error rate, the better the performance.

*F. Parameter Sensitivity*

In this section, the sensitivities of the parameters $\beta, \gamma, \alpha, l, d$ on the performance of DNE-SBP are reported. Figs. 3 and 4 show the parameter sensitivity of DNE-SBP for link sign prediction and community detection, respectively.

*1) Larger Penalty on the Reconstruction Errors of Observed Connections:* $\beta$ denotes the ratio of the penalty on the reconstruction errors of nonzero input elements over that of zero input elements. As shown in Figs. 3(a) and 4(a), $\beta > 1$ leads to much better link sign prediction (i.e., higher AUC score) and also much better community detection (i.e., lower error rate) than $\beta = 1$. This demonstrates that it is highly effective to assign larger penalty to make the SAE more prone to reconstruct the observed links than unknown connections.

*2) Larger Penalty and Stronger Pairwise Constraint on Negative Links:* $\gamma$ specifies the ratio of penalty on the reconstruction errors of negative links over that of positive links, and the ratio of weight of the pairwise constraint targeting for the negative links over that of the positive links. As shown in Fig. 3(b), for link sign prediction, $\gamma > 1$ can achieve much higher AUC scores than $\gamma = 1$. This demonstrates the significant effectiveness and necessity of imposing larger penalty and stronger pairwise constraint on the scarce negative links to handle the highly imbalanced data (i.e., overwhelming positive links) in the real-world signed networks. Also, when $\gamma \leq 3$, a higher value of $\gamma$ would lead to higher AUC score, while after that, the AUC scores will slightly decrease. Note that in the Wiki dataset, the ratio of the number of positive links over that of negative links is 3.63. Thus, it indicates that setting $\gamma = \text{floor}(\sum_{i=1}^{n}\sum_{j=1}^{n} A_{ij}^{+} / \sum_{i=1}^{n}\sum_{j=1}^{n} A_{ij}^{-})$ is reasonable for DNE-SBP to achieve a good performance for link sign prediction. On the other hand, as shown in Fig. 4(b), all different values of $\gamma$ can achieve a satisfactory low error rate for signed network clustering. It indicates that the performance of DNE-SBP for signed network clustering is insensitive to the value of $\gamma$.

*3) Effect of Semisupervised Learning:* $\alpha_k$ denotes the weight of pairwise constraints at the $k$th layer of SAE. As shown in Figs. 3(c) and 4(c), $\alpha_1 > 0$ would contribute to better link sign prediction and also better signed network clustering performance, as compared to $\alpha_1 = 0$. This demonstrates the effectiveness of designing a semisupervised SAE to capture the structural balance property for signed network embedding. In addition, as shown in Fig. 3(d), when $k \geq 2$, the AUC scores of DNE-SBP are insensitive to the value of $\alpha_k$. This indicates that for link sign prediction, incorporating the pairwise constraints to the first layer of SAE is much more effective than doing that for the deeper layers of SAE. However, as shown in Fig. 4(d), when $k \geq 2$, $\alpha_k > 0$ can significantly reduce the error rate achieved by $\alpha_k = 0$. Thus, in contrast to link sign prediction, incorporating pairwise constraints at the deeper layers of SAE is more effective for community detection than doing that at the first layer of SAE.

*4) Effect of Deep Neural Network Architecture:* $l$ denotes the number of layers in the SAE. We constructed five SAEs, with the layer configuration setting as 7118-64, 7118-256-64, 7118-512-256-64, 7118-512-256-128-64, and 7118-1024-512-256-128-64, respectively. The number of layers in these five SAEs are 1, 2, 3, 4, and 5, respectively. Then, for all the five SAEs, we employed the deepest hidden representations as the node vector representations, which are with the same dimensionality as $d = 64$. As shown in Fig. 3(e), $l > 1$ contributes to much higher AUC score than $l = 1$. While when $l > 2$, the AUC score does not further increase as $l$ increases. This reveals that building a 2-layer SAE can contribute to much better link sign prediction performance, as compared to a basic auto-encoder. However, the link sign prediction performance cannot be further improved, even though a deeper SAE is built. In contrast, as shown in Fig. 4(e), for signed network clustering, the error rate would keep decreasing as $l$ increases. This indicates that the deeper SAE framework contributes to better signed network community detection performance. Such interesting differences between link prediction and community detection could be explained by the fact that link prediction generally focuses on the concrete local neighborhood structure, thus more abstract feature representations might not be more informative. However, community detection requires to capture the global network structure which is more abstract, thus the deeper SAE framework would be more powerful for learning the meaningful feature representations.

*5) Effect of Embedding Dimension:* $d$ indicates the dimensionality of the node vector representation learned by the deepest layer of SAE. As shown in Figs. 3(f) and 4(f), when $d \in \{64, 128, 256\}$, DNE-SBP always achieves the good performance for both link sign prediction and signed network clustering.

## V. Conclusion

Network embedding has become a rather hot topic over the past few years, with the goal of learning the low-dimensional feature vector representations which can well preserve the original network structures. Although several promising network embedding algorithms have been proposed recently, the vast majority of them are only designed for unsigned networks, without considering the polarities of links in the signed networks. In this paper, we propose a DNE-SBP model to employ a semisupervised SAE to learn the nonlinear node vector representations for a given signed network, by reconstructing its signed adjacency matrix. To handle the overwhelmingly positive links in the real-world signed networks, we impose larger penalty on the reconstruction errors of negative links to make the SAE focus more on reconstructing the scarce negative links than the abundant positive links. In addition, to capture the extended structural balance property of signed networks, we incorporate the ML and CL pairwise constraints to map the positively connected nodes closer to each other and map the negatively connected nodes more far apart from each other in the embedding space. Based on the low-dimensional node vector representations learned by DNE-SBP, we apply vector-based machine learning techniques to conduct link sign prediction and signed network community detection. Comprehensive experimental results demonstrate that the proposed DNE-SBP model significantly outperforms the state-of-the-art network embedding algorithms for graph representation learning in the signed networks. In addition, we observe that building a 2-layer SAE has contributed to a good performance for link sign prediction. However, building the deeper SAE which learns more abstract feature representations, is more beneficial for community detection.

In the future, we can apply DNE-SBP to recommender systems, by modeling users and items as nodes, and the "like" and "dislike" relations as the "positive" and "negative" links. In addition, by integrating network structure and the rich content information (e.g., users/items features), we can extend DNE-SBP to heterogeneous network embedding [54]. Then, we can make recommendations for users based on the link sign prediction results of DNE-SBP and conduct customer segmentation according to the network clustering results of DNE-SBP.

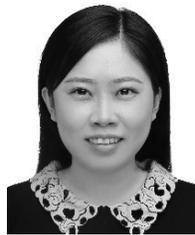

**Xiao Shen** received the double B.Sc. degrees in electronic engineering from the Beijing University of Posts and Telecommunications, Beijing, China, and the Queen Mary University of London, London, U.K. in 2012, and the M.Phil. degree in computer science from the University of Cambridge, Cambridge, U.K., in 2013. She is currently pursuing the Ph.D. degree in computer science with the Department of Computing, Hong Kong Polytechnic University, Hong Kong.

Her current research interests include feature representation learning, deep learning, transfer learning, and data mining in complex networks.

Ms. Shen was a recipient of the Excellent Academic Performance Scholarship Awarded by the Queen Mary University of London and the Hong Kong Ph.D. Fellowship Awarded by the Research Grants Council of the SAR Government.

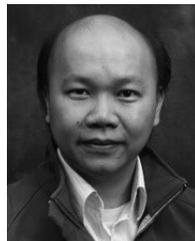

**Fu-Lai Chung** received the B.Sc. degree from the University of Manitoba, Winnipeg, MB, Canada, in 1987 and the M.Phil. and Ph.D. degrees from the Chinese University of Hong Kong, Hong Kong, in 1991 and 1995, respectively.

In 1994, he joined the Department of Computing, Hong Kong Polytechnic University, Hong Kong, where he is currently an Associate Professor. He has published widely in prestige international journals, including the IEEE TRANSACTIONS ON NEURAL NETWORKS AND LEARNING SYSTEMS, the IEEE TRANSACTIONS ON FUZZY SYSTEMS, the IEEE TRANSACTIONS ON CYBERNETICS, the IEEE TRANSACTIONS ON KNOWLEDGE AND DATA ENGINEERING, *Pattern Recognition*, and *Neural Networks*. His current research interests include deep learning, transfer learning, adversarial learning, social network analysis and mining, recommendation systems, and big data learning.

Dr. Chung also serves on program committees of top international conferences, including IEEE ICDM, AAAI, and ICPR.